\documentclass[aps,prb,superscriptaddress,twocolumn]{revtex4}
\usepackage{amsmath,epsfig,color,amssymb}
\usepackage{graphicx}
\usepackage{graphicx}
\usepackage{dcolumn}
\usepackage{bm}
\usepackage{graphicx}
\usepackage{amsmath, amsfonts, amssymb,mathrsfs}
\usepackage{pstricks}
\usepackage{amsxtra}
\usepackage{amsthm}
\usepackage{natbib}

\def\be{\begin{equation}}
\def\ee{\end{equation}}
\def\bea{\begin{eqnarray}}
\def\eea{\end{eqnarray}}
\newcommand{\ket}[1]{\mbox{$|#1\rangle$}}
\newcommand{\bra}[1]{\mbox{$\langle#1|$}}

\def\be{\begin{equation}}      
\def\ee{\end{equation}}
\def\beu{\begin{equation*}}   
\def\eeu{\end{equation*}}

\providecommand{\abs}[1]{\left\lvert#1\right\rvert}   
\DeclareMathOperator{\trace}{Tr}      
\providecommand{\ket}[1]{\left|#1\right\rangle}

\providecommand{\bra}[1]{\left\langle#1\right|}
\providecommand{\mean}[1]{\left\langle#1\right\rangle}

\providecommand{\bq}{{\bm{q}}}

\begin{document}
\title{Probing electron-phonon interactions in the charge-photon dynamics of cavity-coupled  double quantum dots}
\author{M.~J.~Gullans}
\affiliation{Department of Physics, Princeton University, Princeton, New Jersey 08544, USA}
\affiliation{Joint Quantum Institute and Joint Center for Quantum Information and Computer Science, NIST/University of Maryland, College Park, Maryland 20742, USA}
\author{J.~M.~Taylor}
\affiliation{Joint Quantum Institute and Joint Center for Quantum Information and Computer Science, NIST/University of Maryland, College Park, Maryland 20742, USA}
\author{J. R. Petta}
\affiliation{Department of Physics, Princeton University, Princeton, New Jersey 08544, USA}
\begin{abstract}
Electron-phonon coupling is known to play an important role in the charge dynamics of semiconductor quantum dots.  Here we explore its role in the combined charge-photon dynamics of cavity-coupled double quantum dots.  Previous work on these systems has shown that  strong electron-phonon coupling leads to a large contribution to photoemission and gain from phonon-assisted emission and absorption processes.    We compare the effects of this phonon sideband in three commonly investigated gate-defined quantum dot material systems: InAs nanowires and GaAs and Si two-dimensional electron gases (2DEGs).  We compare our theory with existing experimental data from cavity-coupled InAs nanowire and GaAs 2DEG double quantum dots and find quantitative agreement only when the phonon sideband and photoemission processes during lead tunneling are taken into account.  Finally we show that the phonon sideband also leads to a sizable renormalization of the cavity frequency, which allows for  direct spectroscopic probes of the electron-phonon coupling in these systems.  
 \end{abstract}
\maketitle

\section{Introduction}

Lasing and photoemission dynamics serve as a powerful  probe of  light-matter interactions.\cite{CohenBook,PowellBook}  Recent years have seen solid-state masing and photoemission  pushed into the microwave quantum optical limit of few-level systems interacting with single microwave photons.\cite{Girvin09}  These achievements are due to the development of hybrid devices that integrate superconducting cavities with other coherent solid-state quantum systems such as superconducting qubits\cite{Nomura10,Chen14,Cassidy17} or semiconductor quantum dots.\cite{Viennot14,Liu14,Liu15,Stockklauser15}  In both the superconducting qubit and quantum dot systems re-pumping for maser operation is induced by a finite source-drain bias, which results in population inversion.  However, the manner in which each system is coupled to the environment differs significantly.  In particular, in III-V quantum dots, piezoelectric electron-phonon coupling is strong and can lead to inelastic charge relaxation without photon emission,\cite{Fujisawa98,Kouwenhoven01} or a phonon-assisted photoemission process.\cite{Gullans15,Muller17}

At a finer level, there are strong differences even among different semiconductor quantum dot platforms.  In InAs nanowires confinement effects strongly modify the phonon spectrum and resulting electron-phonon coupling.\cite{Weber10,Roulleau11}  These nanowire systems can be contrasted with gate defined quantum dots in GaAs or Si two-dimensional electron gases (2DEGs).  In GaAs 2DEGs the phonon coupling at low energies is dominated by bulk piezoelectric coupling, while in the  Si there is just bulk deformation potential coupling due to the inversion symmetry of the unit cell.\cite{YuBook}  The dimensionality and strength of electron-phonon coupling can thus impact photoemission properties of semiconductor quantum dots.

\begin{figure}[htbp]
\begin{center}
\includegraphics[width = .49 \textwidth]{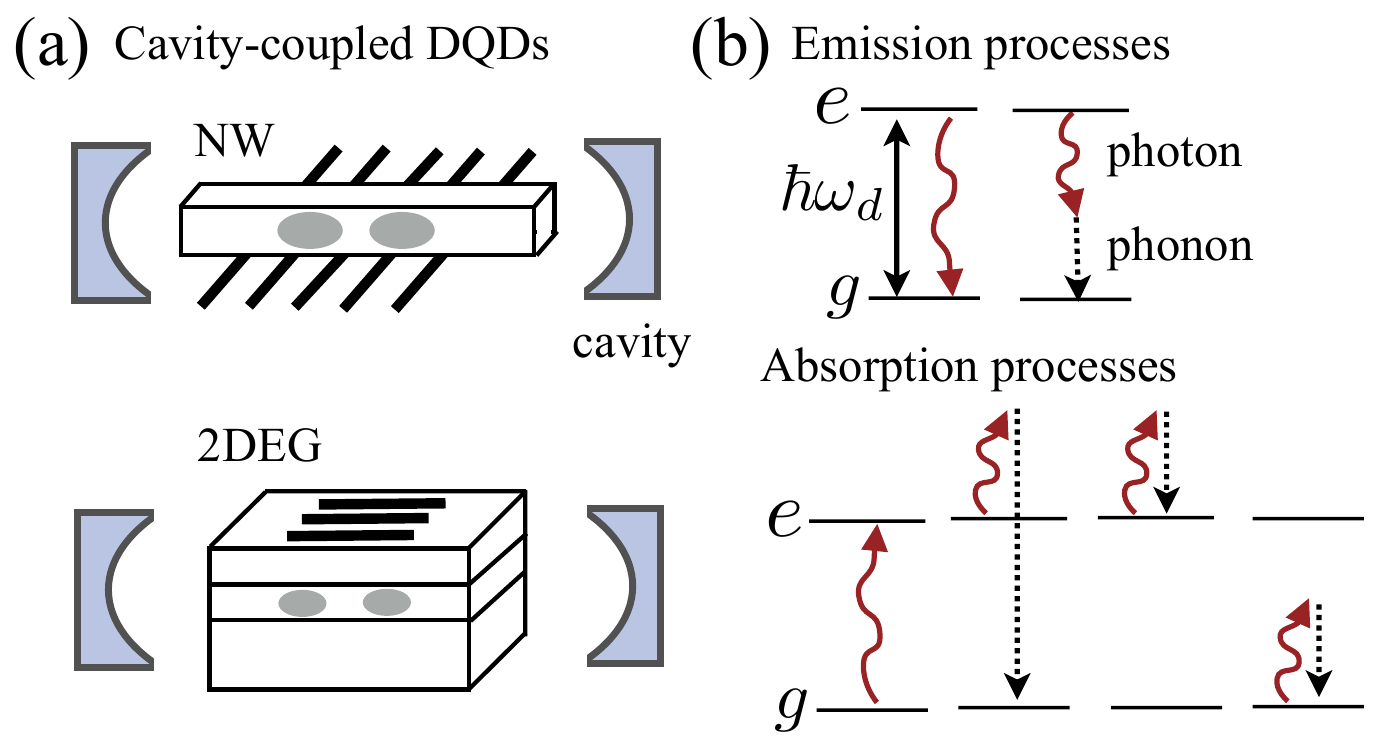}
\caption{(a) Schematic of two types of cavity-coupled  DQDs we consider based on nanowires (NW) or 2DEGs.  Despite the similarities in the low-energy descriptions of the system, the surrounding environment is dramatically different in the two cases.  
Differences in the phonon environment have a strong effect on the charge-photon dynamics due to the presence of phonon-assisted emission and absorption processes. 
(b) The charge states of the DQD form an effective two-level system (TLS).  Direct and phonon-assisted emission and absorption processes at zero temperature are depicted by the arrows.  }
\label{fig:cartoon}
\end{center}
\end{figure}

The purpose of this paper is to quantitatively compare the role of electron-phonon interactions in the photoemission and gain of cavity-coupled double quantum dots (DQDs) in the three material systems illustrated  in Fig.~\ref{fig:cartoon}(a): InAs nanowires and GaAs and Si 2DEGs.  Previous theoretical work identified four phonon-assisted emission and absorption processes [see Fig.~\ref{fig:cartoon}(b)] that play a key role in the charge-photon dynamics of these system.\cite{Gullans15,Muller17}  We find that the primary distinction between these material systems is the strength of the coupling to the phonon sideband.  We find that the phonon sideband plays the largest role in InAs nanowire DQDs due to their enhanced 1D phonon density of states at low-energy.  In GaAs 2DEG DQDs, the phonon sideband plays a weaker role due to the reduction in the low-energy phonon density of states in going from 1D to 3D, while phonons have the weakest effect in Si 2DEG DQDs due to the absence of piezoelectric coupling.  We compare our theory to available experimental data in InAs\cite{Liu15} and GaAs.\cite{Stockklauser15} We conclude by discussing a technique for direct detection of the electron-phonon coupling by measuring shifts in the cavity frequency induced by the phonon sideband.  

The paper is organized as follows.  In Sec.~\ref{sec:dqd} we present our microscopic model for the problem that includes a low-energy description of the coupling of the DQD effective two-level system (TLS) to leads, phonons, and cavity photons.  In Sec.~\ref{sec:ness} we determine the properties of the non-equilbrium steady state (NESS) of the DQD in the presence of finite source-drain bias, which continually re-pumps the excited state.  In Sec.~\ref{sec:photon} we use our model to compute the cavity response to an external drive and the photoemission rate in the absence of a drive in the NESS.  We then compare these predictions across the three different material systems mentioned above, as well as to experimental data.  Throughout this work we restrict our discussion to the regime below the masing threshold. In Sec.~\ref{sec:phonon} we show how the phase response of the cavity serves as a detailed spectroscopic probe of the electron-phonon coupling in this system, as demonstrated in a recent experiment on a suspended InAs nanowire DQD.\cite{Hartke17} 

\section{Model}
\label{sec:dqd}
Due the the large charging energy of each quantum dot ($ E_c\sim 5$~meV), we model the system by including just three DQD charge states $\ket{0}$, $\ket{L}$, $\ket{R}$, where $\ket{0}$ has $(M,N)$ electrons, $\ket{L}$ has $(M+1,N)$ electrons, and $\ket{R}$ has $(M,N+1)$ electrons  in the (left, right) dots.  The low-energy Hamiltonian describing the coupling between these charge states, the single mode of a microwave cavity, the leads, and lattice phonons  is given by
 \begin{align} \label{eqn:H0}
 H&=H_0+H_\ell + H_{ep},\\
 H_0 & =   \frac{\epsilon}{2} \sigma_z + t_c \sigma_x + \hbar \omega_c a^\dagger a + \hbar g_c \sigma_z (a+a^\dagger), \\
 H_\ell& = \sum_k (\epsilon_{k} - \mu_\ell)d_{k \ell}^\dagger d_{k \ell} + (\epsilon_k - \mu_r)d_{k r}^\dagger d_{k r} \\ \nonumber
 &+ \sum_k \big( t_{k\ell} d_{k\ell}^\dagger c_{\ell} + t_{kr} d_{k r}^\dagger c_{r} + h.c.),\\
 H_{ep}&=  \sum_{q,\nu}  \hbar \omega_{q\nu}a_{q\nu}^\dagger a_{q\nu}+ \hbar \lambda_{q \nu} \sigma_z (a_{q\nu}^\dagger+a_{q\nu}),
\end{align}
where $\sigma_\mu$ are Pauli-matrices operating in the orbital subspace $\ket{L}$ and $\ket{R}$, $\epsilon$ is the detuning between the two dots, $t_c/\hbar$ is the interdot tunneling rate, $\omega_c$ is the bare cavity frequency, $g_c$ is the DQD-cavity coupling,  and $a^\dagger\,(a)$ are cavity photon creation (annihilation) operators.   $H_\ell$ describes the coupling to the leads, where $d_{k\ell(r)}$ are the Fermion  annihilation operators for the left(right) lead, $\epsilon_k$ is the electron dispersion in the leads, $\mu_{\ell(r)}$ is the chemical potential in the left(right) lead, $c_{\ell(r)}$ is the Fermion annihilation operator for the left(right) dot, and $t_{k \ell(r)}$ are tunneling matrix elements between the leads and the dots. In the electron-phonon interaction $H_{ep}$,
 $\omega_{q\nu}$ is the phonon dispersion, $\lambda_{q \nu}$ is a coupling constant that depends on momentum $q$ and mode index $\nu$ of the phonons and $a_{q\nu}^\dagger(a_{q\nu})$ are phonon  creation~(annihilation) operators.  

\section{Non-Equilibrium Steady State of the DQD}
\label{sec:ness}
In this section we integrate out the leads to zeroth order in the cavity-DQD coupling to derive a master equation describing the three charge states of the DQD  under application of a source-drain bias $V_{sd}=\mu_\ell -\mu_r$.   We find the non-equilibrium steady-state (NESS) of the DQD and response functions, which are later used to determine the steady-state cavity response perturbatively in the cavity-DQD coupling.

To solve the dynamics of this model with a finite  $V_{sd}$, we transform to the eigenstates of $H_0$, for $\epsilon>0$
 \begin{align}
\ket{e} = \cos (\theta/2) \ket{L}-  \sin (\theta/2) \ket{R}, \\
\ket{g}= \sin (\theta/2) \ket{L}+  \cos (\theta/2) \ket{R}, 
\end{align}
where $\ket{g/e}$ are the ground/excited energy states of the DQD with energy $\hbar \omega_\pm= \pm  \sqrt{\epsilon^2/4+t_c^2}$, $\hbar \omega_d=\hbar(\omega_+-\omega_-)$ is the energy splitting between these two states, and $\theta = \tan^{-1}(2t_c/\epsilon)$.

To model the current flow in the absence of the cavity we integrate out both the phonons and the electrons in the leads in a Born-Markov approximation to arrive at the master equation for three DQD states $(\epsilon>0)$

\be \label{eqn:rho1}
\begin{split}
\dot{\rho}& = -i [H_0,\rho] + \Big( \alpha \Gamma_L \mathcal{D}[\ket{e} \bra{0}]  + \alpha \Gamma_R  \mathcal{D}[\ket{0}\bra{g}]  \\
&+(1-\alpha) \Gamma_L \mathcal{D}[\ket{g}\bra{0}] +(1-\alpha) \Gamma_R \mathcal{D}[\ket{0} \bra{e}]\Big) \rho  \\
&+ \gamma_d \big[ (1+n_d) \mathcal{D}[\ket{g}\bra{e}] +n_d \mathcal{D}[\ket{e}\bra{g}] \big] \rho ,
\end{split}
\ee

where $\alpha=\cos^2 (\theta/2)$ and  the Lindblad superoperators act according to $\mathcal{D}[A]\rho=-1/2 \{ A^\dagger A,\rho \} + A \rho A^\dagger$ for any operator $A$.    For $\epsilon<0$, the $\ket{g/e}$ states in the four lead tunneling terms in Eq.~(\ref{eqn:rho1}) are reversed.  
The relaxation rates that enter this master equation are the  energy dependent tunneling rates from the left lead onto the left dot ($\Gamma_L$) and from right dot into the right lead ($\Gamma_R$):
\begin{align}
\Gamma_{L}(\nu)&=2\pi/\hbar \sum_k \abs{t_{k\ell}}^2 f_\ell( \hbar \nu )  \delta(\hbar \nu - \epsilon_k),\\
\Gamma_{R}(\nu)&=2\pi/\hbar \sum_k \abs{t_{kr}}^2 [1-f_r(\hbar \nu )]  \delta(\hbar \nu - \epsilon_k),
\end{align}
 where $f_{\ell,r}(\varepsilon) = [e^{(\varepsilon- \mu_{\ell,r})/k_B T}+1]^{-1}$ is the Ferm-Dirac distribution function for the left/right lead.  When $ V_{\rm sd} \gg \hbar  \omega_d$ the system undergoes sequential single-electron tunneling events within a finite bias triangle in the DQD charge-stability diagram.\cite{Kouwenhoven01}  In this work, we assume $\Gamma_{L,R}$ are independent of energy over the relevant range of $\epsilon$ and $t_c$.    The phonon induced inter-dot charge relaxation rate is given by 
 \begin{align} \label{eqn:gamd}
 \gamma_d(\epsilon)& = 2 \pi \sin^2 \theta \, J(\omega_d), \\ \label{eqn:Jph}
 J(\nu) &= \sum_{q,\nu} |\lambda_{q \nu}|^2 \delta[\nu - \omega_{q\nu}],
 \end{align}
  where $J(\nu)$ is the phonon spectral density (see Appendix A). 

The NESS of the DQD $\rho_{ss}$ is a diagonal density matrix. The steady-state populations in the ground $p_g$ and excited $p_e$ states of the DQD can be found analytically from Eq.~(\ref{eqn:rho1}).  The steady state current through the DQD for $\epsilon>0$ is
\be \label{eqn:Ip}
\begin{split}
I &= e \Gamma_R \trace(\rho_{ss} \ket{R} \bra{R}) =e\Gamma_R [(1-\alpha) p_e + \alpha p_g],
\end{split}
\ee
where $e$ is the electronic charge and, for $\epsilon<0$,
\be \label{eqn:Im}
I=e \Gamma_R [(1-\alpha) p_g+ \alpha p_e ].
\ee
In addition to the NESS, the master equation can be used to find all steady-state DQD correlation functions $\mean{\prod_i \sigma_{\nu_i}(t_i)}$ to zeroth order in $g_c$  using the quantum regression theorem. \cite{QuantumOpticsBook}

\section{Photon Emission and Gain}
\label{sec:photon}

In the presence of a finite source-drain bias, the resulting current drives a steady-population in the excited state of the effective two-level system (TLS) of the DQD.  Repopulation of this  excited state leads to a continuous rate of photon emission into the cavity.  When the population in the excited state exceeds the population in the ground state, the cavity-DQD system is effectively inverted.\cite{Childress04,Jin12,Kulkarni14}  Population inversion will lead to a gain response to an input cavity field.  These two effects, photon emission and gain, are associated with different correlation functions of the driven cavity-coupled DQD.  Experiments are able to probe both effects in these systems.\cite{Liu14,Viennot14,Stockklauser15}  In this section, we calculate   the gain of the system when driven by an external cavity field and the steady state photon emission rate in the absence of a cavity drive.  We compare our results between the different material systems mentioned in the introduction, as well as recent experimental data.\cite{Liu15,Stockklauser15} In all three material systems we find that the phonon sideband plays an important role in understanding and modeling the resulting cavity correlation functions.

 To find the cavity response we return to the original Hamiltonian and derive Heisenberg-Langevin equations for the cavity operators.  Unlike the description of current through the DQD, this approach is typically more tractable than solving  for the full density matrix of the cavity-field due to the infinite Hilbert space of the cavity.  We first regroup $H$ into a system Hamiltonian for the DQD and cavity, a reservoir describing the bare phonons, leads, and cavity environment,  and the coupling between them 
 \begin{align} \label{eqn:H}
H&=H_{S}+ H_{R}+H_{SR}, \\
H_S &= \frac{\omega_d}{2} \sigma_z +\omega_c a^\dagger a+ g_c \sin \theta (a^\dagger \sigma_-+ h.c.), \\
H_R&=\sum_k \big[ (\epsilon_{k} - \mu_\ell)d_{k \ell}^\dagger d_{k \ell} + (\epsilon_k - \mu_r)d_{k r}^\dagger d_{k r}\big] \\ \nonumber&+  \sum_{q,\nu} \omega_{q\nu} a^\dagger_{q\nu}a_{q\nu} + \sum_n \omega_n b^\dagger_n b_n, \\
H_{SR}& =-R_c^\dagger\, a  - R_{ L}^\dagger c_\ell - R_{ R}^\dagger c_r - R_d^\dagger\, \sigma_-  \\ \nonumber
&- R_\phi^\dagger\, a \sigma_z- R_e^\dagger\, a^\dagger \sigma_- - R_a^\dagger\, a \sigma_-   + h.c.,
\end{align}
where we introduced a bath of modes that couple to the cavity field $b_n$ and neglected counter-rotating terms and higher order terms in $(g_c,\lambda_{q,\nu})$ in $H_S$ and $H_R$.  The reservoir operators are given by
\begin{align}
R_c &= \sum_n \tau_n b_n,~~R_{L (R)} = - \sum_k  t_{k\ell(r)} d_{k\ell(r)} , \\ \label{eqn:rd}
R_d& =-\frac{2 t_c}{\omega_d}  \sum_{q,\nu} \lambda_{q \nu} a_{q\nu}, ~R_\phi = \frac{8 t_c^2}{\omega_d} \sum_{q,\nu} \frac{ g_c \lambda_{q \nu}}{\omega_d^2-\omega_c^2} a_{q\nu}, \\
R_e&= \frac{4 t_c \epsilon}{\omega_d^2} \sum_{q,\nu} \frac{ g_c \lambda_{q \nu}}{\omega_c \omega_{q\nu}} (\omega_{q\nu}+\omega_c) a_{q\nu}, \\
R_a &=\frac{4 t_c \epsilon}{\omega_d^2} \sum_{q,\nu} \frac{ g_c \lambda_{q \nu}}{\omega_c \omega_{q\nu}} (\omega_{q\nu}-\omega_c) a_{q\nu}. \label{eqn:ra}
\end{align}
$R_{\phi,e,a}$ are the reservoir operators for the phonon sideband processes, which were derived in Refs.~\onlinecite{Gullans15,Muller17} and account for phonon-assisted emission and absorption of the cavity field [see Fig.~\ref{fig:cartoon}(b)].  To avoid infrared singularities in the denominators of Eqs.~(\ref{eqn:rd})--(\ref{eqn:ra}) we regularize the  phonon propagator at low energies.  Our regularization procedure is described below.  In principal, such effects could be accounted for self-consistently in our theory using the diagrammatic approach detailed in Ref.~\onlinecite{Muller17}; however, it is likely that a proper microscopic treatment of these effects requires a careful consideration of the entire phonon environment, which is beyond the scope of the present work.  Furthermore, we find that all physical observables we compute are independent of this regularization procedure except in the case of InAs nanowires in the restricted region  $\omega_d \approx \omega_c$, where there is an explicit dependence on the infrared cutoff.  

We can express the reservoir operators in terms of Langevin noise operators to arrive at the Heisenberg-Langevin equations for the cavity field and charge  \cite{QuantumOpticsBook2}
\begin{align} \label{eqn:a}
\dot{a} &= -[ (\kappa - {\chi})/2+i (\delta  + \delta_R) ] a + i g_c \sin \theta \sigma_- + \Omega  \\ \nonumber
&+ \sigma_- \mathcal{F}_e^\dagger + \sigma_+ \mathcal{F}_a  + \sigma_z \mathcal{F}_\phi+ \mathcal{F}_c, \\ 
\label{eqn:chi}
{\chi}&= r_c\, \mean{\sigma_z} + (\gamma_e-\gamma_a)p_e - \gamma_{e}' p_g - \gamma_\phi (p_g+p_e) \\\nonumber
& + (\gamma_e n_e +\gamma_e'  n_e'  -  \gamma_a n_a  )\mean{\sigma_z}  - \kappa_R ,\\
r_c&= \frac{ 2 g_c^2 \sin^2 \theta}{\Gamma^2 + \Delta^2} \Gamma,~~ \Gamma = \gamma_d  (n_d+1/2) +\Gamma_R/2, \\
n_e &= n_p(\Delta),~~n_e'=n_p(-\Delta),\\
n_a&=n_p(\omega_d +\omega_c),~~n_\phi=n_p(\omega_c),
\end{align}
where we have made the additional approximation of treating the DQD operators in mean field theory when evaluating $\chi$, $\delta =\omega_c-\omega$ is the detuning between the bare cavity frequency and the drive frequency $\omega$, $\delta_R \propto g_c^2$ is the renormalization of the cavity frequency due to interactions with the quantum dot (see Sec.~\ref{sec:phonon} for a more detailed discussion of this term), $\Omega$ is the drive amplitude, $\Delta = \omega_d - \omega_c-\delta_R$, $\kappa$ is the cavity decay rate,  $\chi$ is the mean-field gain rate and $r_c$ is the direct photoemission and absorption rate of the TLS.   The noise operators $\mathcal{F}_\nu$ are associated with each of the reservoir fields and, in the Markov approximation, satisfy
\begin{align}
[\mathcal{F}_\nu(t'),\mathcal{F}_\nu^\dagger(t)] &= \gamma_\nu \delta_{\nu\nu'} \delta(t-t'),\\
\langle {\mathcal{F}_{\nu'}^\dagger(t') \mathcal{F}_\nu(t)} \rangle &= \gamma_\nu n_\nu \delta_{\nu\nu'}\delta(t-t'),
\end{align}
where $\gamma_\nu$ is the associated decay rate and $n_\nu$ is the steady-state occupation of the reservoirs, which are assumed to be thermal.  The phonon-assisted emission and absorption terms have the explicit expressions \cite{Gullans15,Muller17}
\begin{align} \label{eqn:game}
\gamma_e&=\frac{ 8 \pi g_c^2 \epsilon^2 \sin^2 \theta  }{\omega_c^2 (\Delta^2+\eta^2)} J(\Delta),~~\gamma_e'=\gamma_e(-\Delta), \\
\gamma_a&=\frac{ 8 \pi g_c^2 \epsilon^2 \sin^2 \theta  }{\omega_c^2 (\omega_d +\omega_c)^2} J(\omega_d +\omega_c),\\ \label{eqn:deph}
\gamma_\phi& = \frac{8 \pi g_c^2 \omega_d^2 \sin^4\theta }{(\Delta^2+\eta^2)(\omega_d+\omega_c)^2} J(\omega_c) ,
\end{align}
where $\gamma_e'$ is a phonon-assisted absorption process from the ground state (not shown in Fig.~\ref{fig:cartoon}) that becomes relevant when $\hbar \omega_c > 2 t_c$.  In the expressions for $\gamma_{e,\phi}$ we regularized any potential infrared divergences that can arise when $ \omega_d$ is equal to the renormalized cavity frequency $\omega_c +\delta_R$ by introducing the renormalization parameter $\eta$.  The physical scale of this parameter is set by the finite lifetime of the phonons and excited state of the DQD.

\begin{figure*}[htb]
\begin{center}
\includegraphics[width = .9 \textwidth]{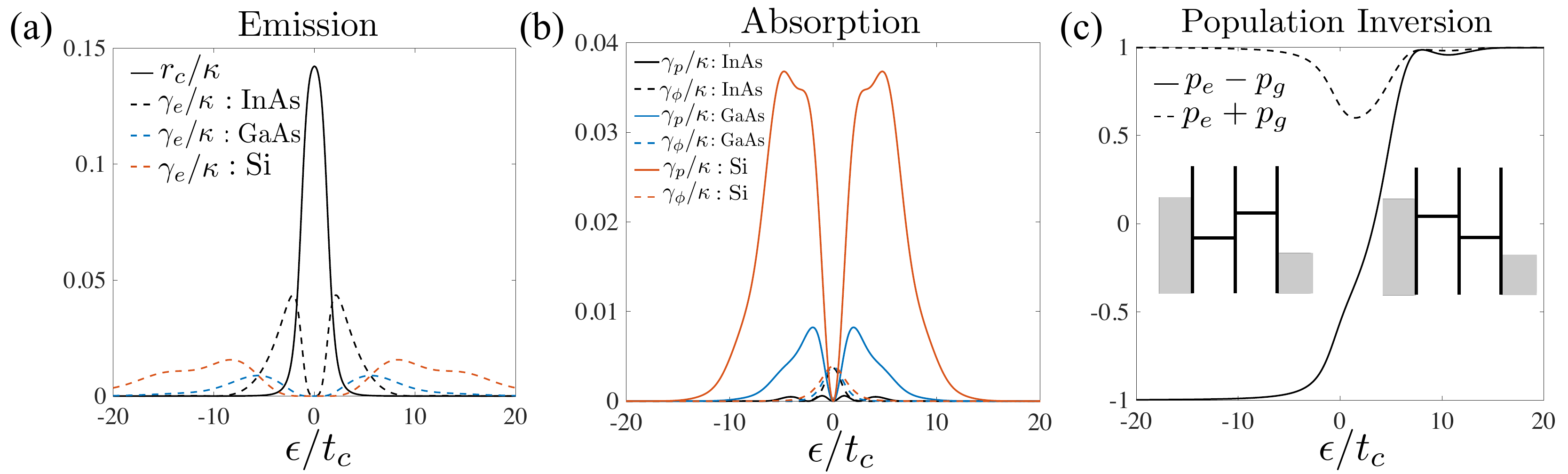}
\caption{(a) Emission and (b) absorption rates for different processes from Fig.~\ref{fig:cartoon}(b--c) as a function of $\epsilon$ for three different material systems considered in this work. (c) Dependence of DQD populations on $\epsilon$ for InAs (results are similar for GaAs and Si).  Large population inversion is only achieved for $\epsilon \gtrsim t_c$ due to asymmetry in level structure with respect to the source-drain bias (inset).  Here we took $g_c/2 \pi =40~$MHz, $\kappa/2\pi = 2$~MHz, $\omega_c/2\pi = 8$~GHz, $\gamma_d(0)/2\pi=1$~GHz, $\omega_0/2\pi = 30~$GHz, $d=120$~nm, $c_n=3\, 000$~m/s,~$c_b=4\, 000$~m/s, $r=0.5$, $ 2t_c/h=8.1$~GHz, $\Gamma_{L,R}/2\pi = 1$~GHz, and $T=100$~mK.}
\label{fig:params}
\end{center}
\end{figure*}

Throughout this work we will be focused on the regime $\chi < \kappa$ below the threshold for masing.
In the presence of an external cavity drive in this regime, the output from the cavity will be in a coherent state with amplitude
\be
\mean{a} = \frac{\Omega}{(\kappa-\chi)/2 + i (\delta+\delta_R)}.
\ee
This expression  shows that the cavity response is strongly sensitive to the photon emission and absorption processes in the system through $\chi$, as well as the renormalization of the cavity frequency $\delta_R$.  We define the normalized gain $G$ and phase response $\phi$ as 
\begin{align} \label{eqn:G}
G(\epsilon,\delta)&= \frac{\kappa^2}{[\kappa-\chi(\epsilon)]^2 + 4 [\delta+\delta_R(\epsilon)]^2}, \\ \label{eqn:phi}
\phi(\epsilon,\delta) & = - \tan^{-1}\bigg( \frac{\delta +\delta_R(\epsilon)}{[\kappa-\chi(\epsilon)]/2} \bigg).
\end{align}

In addition to the linear response of the  cavity, direct photoemission can  be measured in the absence of an input cavity field.    Such measurements probe the two-field correlation functions
\begin{align}   \label{eqn:adaga}
\langle a^\dagger(t+\tau)a(t)\rangle,
\end{align}

which can be expressed directly in terms of the DQD and phonon correlation functions by  integrating Eq.~(\ref{eqn:a}) and applying the quantum regression theorem.\cite{QuantumOpticsBook}
Below threshold and for small values of $g_c$, there is negligible backaction of the cavity on the DQD and, as a result, the DQD correlation functions can be found from the master equation in Sec.~\ref{sec:dqd} [Eq.~(\ref{eqn:rho1})].  The only two-time correlation function that is needed to compute Eq.~(\ref{eqn:adaga}) is 
\begin{align}
\mean{\sigma_+(t+\tau)\sigma_-(t)} &= p_e e^{-(\gamma_\phi+i \Delta)\tau}, \\
\mean{\sigma_+(t)\sigma_-(t+\tau)}& = p_e e^{-(\gamma_\phi - i \Delta)\tau},
\end{align}
where $\tau>0$.
  From this analysis we find the mean photon number in the cavity below threshold can be decomposed into a contribution from the direct interaction with the TLS and the phonon sideband
\begin{align} \label{eqn:adaga2}
n&=\langle {a^\dagger a} \rangle= \langle a^\dag a \rangle_0+ \langle a^\dag a \rangle_{\rm ph}, \\
\langle a^\dag a \rangle_0 &= \frac{r_c p_e}{\kappa-\chi}  ,\\
\langle a^\dag a \rangle_{\rm ph}&= \frac{\gamma_e (n_e+1) p_e}{\kappa-\chi}  +\frac{\gamma_e' n_e' p_e}{\kappa-\chi} + \frac{\gamma_a n_a p_g}{\kappa-\chi}\\
&\nonumber  +\frac{\gamma_\phi n_\phi (p_e+p_g)}{\kappa-\chi}   .
\end{align}
Both Eq.~(\ref{eqn:G})--(\ref{eqn:phi}) and Eq.~(\ref{eqn:adaga2}) have sizable contributions from direct emission and absorption as well as phonon-assisted processes.  Furthermore, they have a complicated dependence on the control parameters of the TLS: $\epsilon$ and $t_c$.  

To better understand the parametric dependence of the cavity response, in Fig.~\ref{fig:params} we  isolate the dependence of each variable as function of the detuning parameter $\epsilon$.  For simplicity we consider the case where the TLS is near resonance with the cavity at $\epsilon=0$, i.e., $\hbar \omega_c \approx 2 t_c$.  We are particularly interested in making a direct comparison between the behavior of the cavity response in the InAs nanowire DQDs as compared to GaAs or Si 2DEG DQDS.  This comparison is achieved by choosing different effective models for the phonon spectral density given in Eq.~(\ref{eqn:Jph}).  For the nanowire, the dominant contribution to $J(\nu)$ at low-frequencies is from piezoelectric coupling to the lowest order phonon mode of the nanowire, which, for small $|q|$, has the dispersion relation $\omega(q)=c_n |q|$.  As derived in a toy model in Appendix A, the phonon spectral density takes the form \cite{Brandes05,Weber10}
\be \label{eqn:Jnw}
J_{\rm nw}(\nu)=  \frac{J_0 d}{c_n \nu} \sin^2(\nu d/c_n) e^{-\nu^2/2\omega_0^2} + J_b(\nu),
\ee
where $J_0$ is a constant scale factor, $d$ is the spacing between dots, and $\omega_0=c_n/a$ is a cutoff frequency.  However, we treat $\omega_0$ as a free parameter in our calculations.  The term $J_b(\nu)$ is a background phonon spectral density to account for contributions from other phonon modes.  In our calculations we take  $J_b(\nu)=r\, J_3(\nu)$ defined in Eq.~(\ref{eqn:J2d}) below with $r$ a free parameter.  
The spectral density at low-frequencies in a conventional 2DEG in type-III or type-IV semiconductors will be dominated by coupling to bulk acoustic phonons. We take a spherically symmetric linear dispersion $\omega(\bq) = c_b |\bq|$ to approximate
\be \label{eqn:J2d}
J_{\alpha}(\nu) =J_0 \left(\frac{ c_n   \nu}{d}\right)^{\alpha-2} \, [1-\mathrm{sinc}(\nu d/c_b)] e^{-\nu^2/2\omega_c^2},
\ee
where $\alpha$ is determined by whether the piezoelectric coupling dominates ($\alpha=3$), as in the case of GaAs 2DEG, or the deformation potential coupling dominates ($\alpha=5$), as in the case of  Si.  Throughout this work we obtain results for $G(\epsilon,\delta),~\phi(\epsilon,\delta)$ and $n$ by taking $J(\nu)=J_{\rm nw}(\nu)$, $J(\nu)=J_3(\nu)$ and  $J(\nu)=J_5(\nu)$ for   InAs nanowires, GaAs 2DEGs, and for Si 2DEGs, respectively.

At low frequencies $J_{\rm nw} \sim \nu$, while $J_{\alpha}(\nu) \sim \nu^{\alpha}$.  From Eq.~(\ref{eqn:game}) we can see that this low-frequency behavior will strongly enhance the phonon-assisted emission process in Fig.~\ref{fig:cartoon}(b) in the nanowire system.  This is clearly observed in Fig.~\ref{fig:params}(a), where we can see that, all other parameters being equal, the phonon-assisted emission in InAs nanowires is much larger than in GaAs and Si.     Furthermore, in the case of GaAs and Si, we see from Fig.~\ref{fig:params}(b) that the rate for phonon-assisted emission is comparable in magnitude to phonon-assisted absorption, implying that the overall contribution of phonons to the gain will be reduced.  In Fig.~\ref{fig:params}(c) we also show the behavior of the population inversion as a function of $\epsilon$.  The asymmetry between positive and negative $\epsilon$ can be understood intuitively because the position of the ground and excited states with respect to the lead are inverted as $\epsilon$ changes sign (see inset).  Since the phonon-assisted emission processes rely on a large population in the excited state, the NESS will only probe these emission processes at positive $\epsilon$. In the regime near $\epsilon \sim t_c$, it is also important to note that there can be substantial population of the $\ket{0}$ state in the NESS when the inter-dot tunneling rate becomes comparable to the lead tunneling rates.  This is seen in Fig.~\ref{fig:params}(c) as a suppression of the total population on the dot $p_g+p_e$ when $\epsilon \sim t_c$.  

\begin{figure*}[htbp]
\begin{center}
\includegraphics[width = .9 \textwidth]{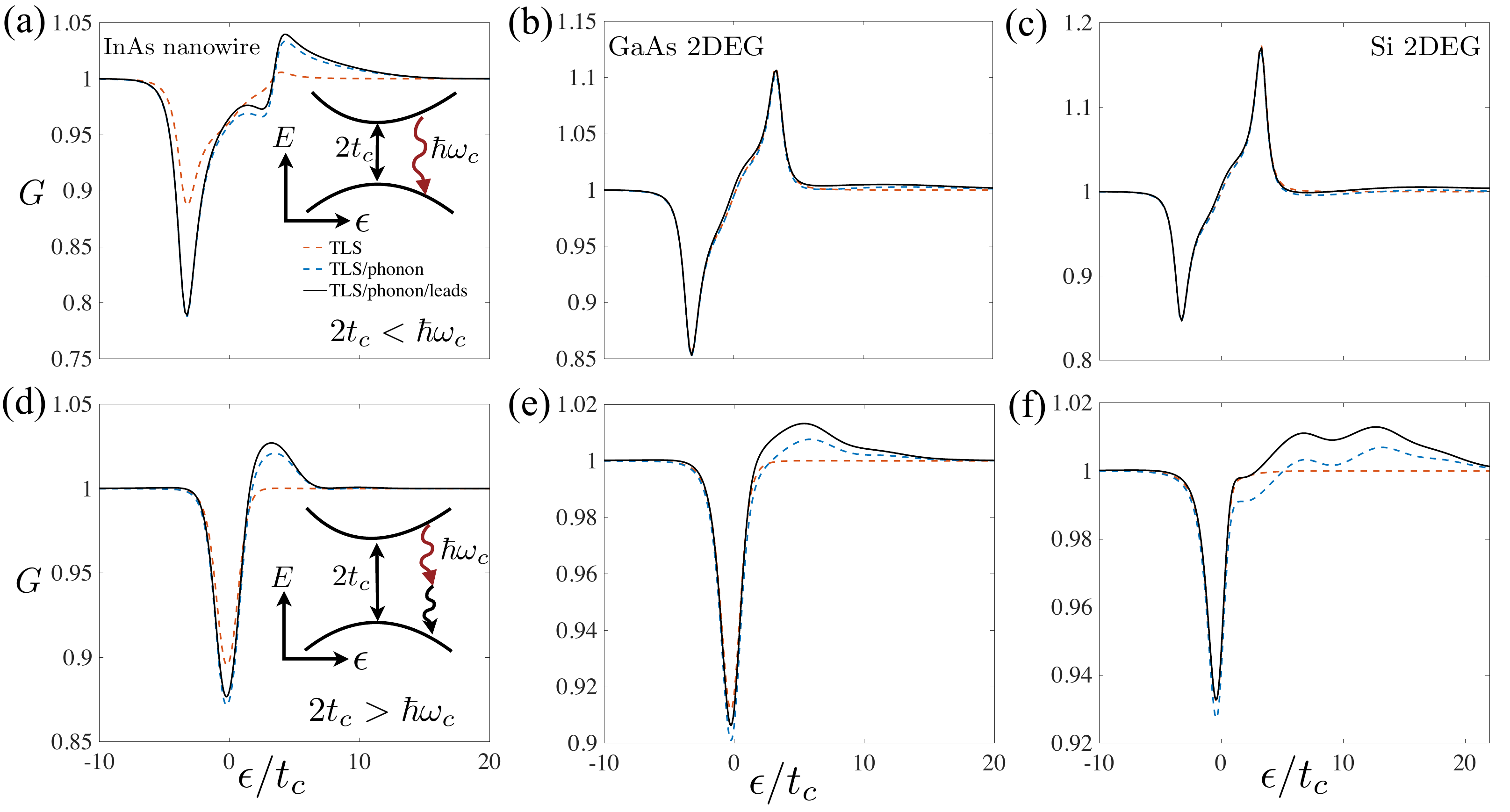}
\caption{(a--c) Normalized gain $G$ in the NESS below threshold as a function detuning for three different material systems in the regime $2 t_c < \hbar \omega_c$. (Red dashed) Effective two-level system (TLS) approximation for the DQD. (Blue dashed) We also include contributions from phonon sideband and (solid) photoemission during lead tunneling. Parameters are as in Fig.~\ref{fig:params} with $2t_c/h=4~$GHz.  Here we also include a small contribution from photoemission during lead emission ($\alpha_{L}=\alpha_R=10^{-5}$).\cite{Liu17} (d--f) Similar to (a--c), but with $2t_c/\hbar =16~$GHz$~ > \hbar \omega_c$.  We see that the relative strength of the gain is comparable in the two cases for the InAs nanowire, but is substantially reduced when $2 t_c > \hbar \omega_c$ for the 2DEGs. }
\label{fig:gain}
\end{center}
\end{figure*}

\begin{figure*}[htbp]
\begin{center}
\includegraphics[width = .9 \textwidth]{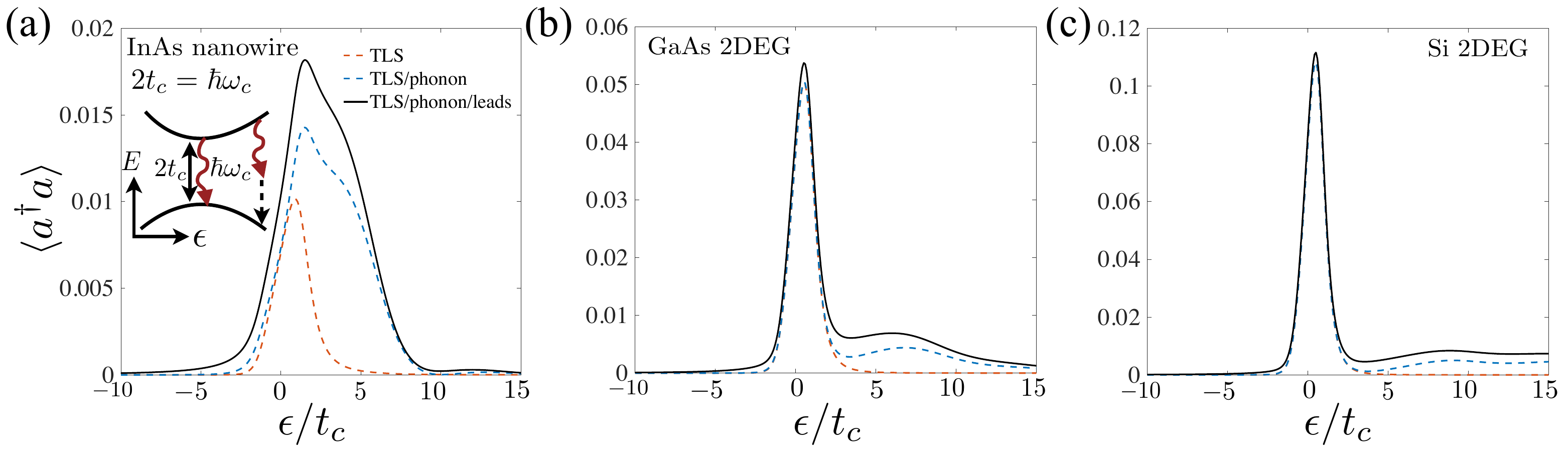}
\caption{(a--c) Mean photon number in the cavity in the NESS below threshold for three different material systems in the regime of maximal photoemission rate $(2 t_c =\hbar \omega_c)$.  Other parameters are as in Fig.~\ref{fig:gain}. We see that for InAs nanowires, the phonon sideband gives a sizable correction to the photoemission near $\epsilon =0$, whereas for the 2DEG the photoemission is dominated by direct photoemission processes near $\epsilon=0$ with a smaller tail at large detuning due to the phonon sideband.}
\label{fig:aa}
\end{center}
\end{figure*}

Figures \ref{fig:gain} and \ref{fig:aa} show the gain and mean photon number in the NESS for the three material systems operating below threshold.  In Figs.~\ref{fig:gain}(a--c) we compare the gain in the regime $2 t_c < \hbar \omega_c$.  In this regime, the point where the cavity is on resonance with the TLS transition energy also overlaps with the region of large population inversion.  This leads to sizable gain in all three material systems, largely independent of the details of the phonon sideband.  Although this regime can be challenging to reach experimentally due to the small tunnel couplings involved and the resulting low-current rates, achieving it provides a material independent route towards masing ($\kappa < \chi$) in these cavity-coupled DQDs.  The opposite regime ($2 t_c> \hbar \omega_c$) has been explored extensively in InAs nanowires where it was shown to allow sizable gain and masing.\cite{Liu14,Liu15,Liu17}  In our previous theoretical work  we attributed the sizable gain in this system to the important role played by the phonon sideband.\cite{Gullans15}  A direct comparison of the InAs nanowire DQDs to  2DEG DQDs in GaAs and Si is presented in Fig.~\ref{fig:gain}(d--f).  For the nanowire, we see that the total amount of gain remains similar to Fig.~\ref{fig:gain}(a);  however, for GaAs and Si 2DEGs, the total amount of gain drops significantly due to the weaker phonon sideband in these systems.  In these calculations we also account for a small correction to $\chi$ and $n$ arising from  photoemission events during lead tunneling\cite{Liu17}
\begin{align} \label{eqn:lead1}
\chi &\to \chi + ( \alpha_{L}+\alpha_R) I/e,\\ \label{eqn:lead2}
n &\to n+\frac{ ( \alpha_{L}+\alpha_R) }{\kappa - \chi} I/e.
\end{align}
Here $I$ is given by Eqs.~(\ref{eqn:Ip})--(\ref{eqn:Im}) and $\alpha_{L/R}$ is the fraction of tunneling events through the left/right barrier that result in a photoemission event.  In Ref.~\onlinecite{Liu17}, detailed theoretical modeling of the threshold dynamics of a single-DQD maser extracted a value of $\alpha_{L,R}$ in the range of $10^{-5}-10^{-4}$ for two separate InAs nanowire devices.   Second-order perturbation theory in $t_{k \mu}$ and $g_c$ for our model Hamiltonian in Sec.~\ref{sec:dqd} predicts the scaling
\be
\alpha_{\mu} \sim  g_c^2\, /\omega_c^2,
\ee
which is  consistent with the order of magnitude of the experimentally extracted values. We leave a more quantitative estimate of these effects for future work.

In Fig.~\ref{fig:aa} we compare the mean photon number in the NESS in the regime of large photoemission $2 t_c = \hbar \omega_c$ for the three different material systems.  Consistent with the results for the gain, we find that the photoemission for InAs nanowires has a large contribution from the phonon sideband.  
  For the 2DEG systems, the phonon sideband plays a smaller role, but still provides a sizable contribution at large positive detunings.  

\begin{figure}[t]
\begin{center}
\includegraphics[width = .3 \textwidth]{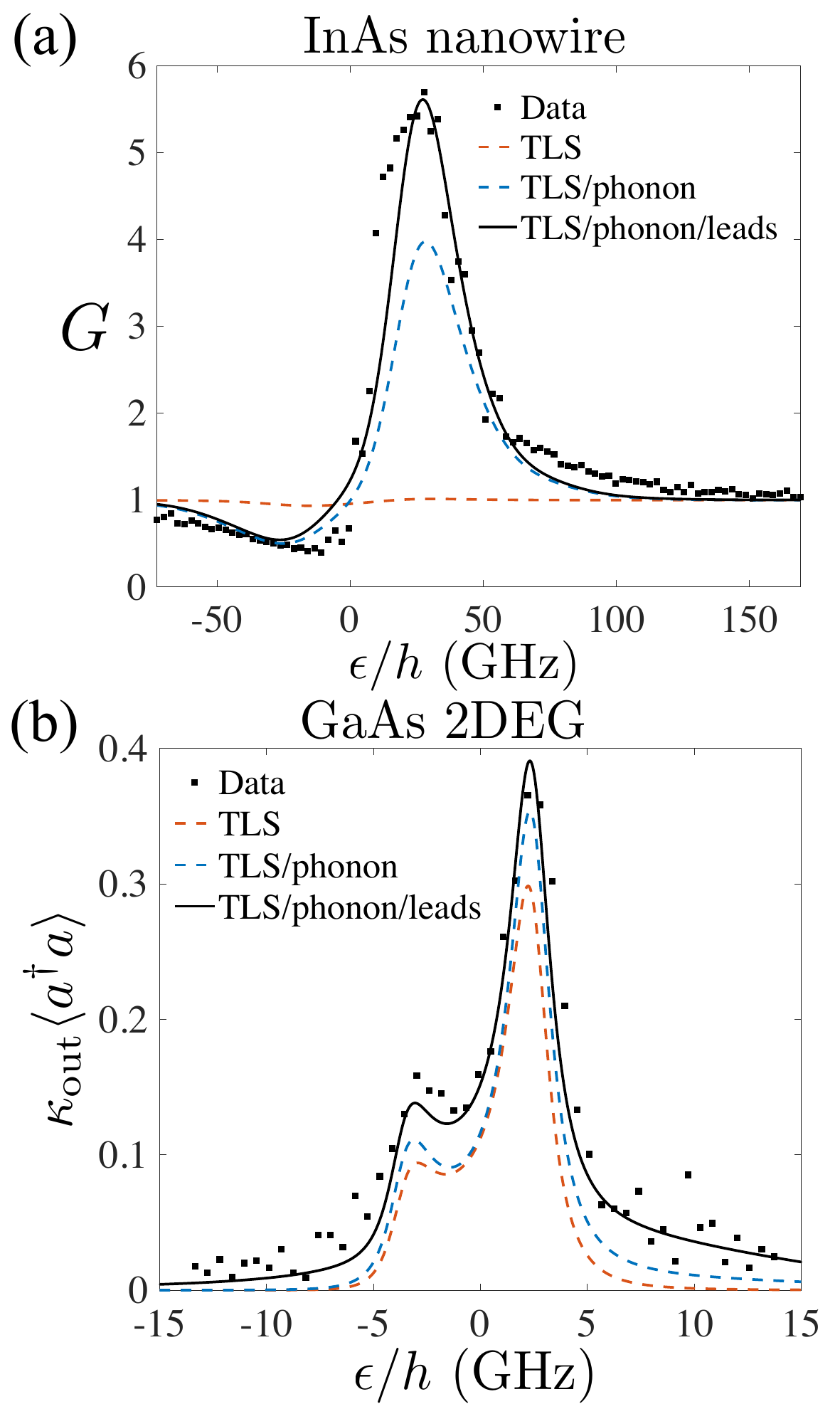}
\caption{(a) Comparison of theory to experimental data from Ref.~\onlinecite{Liu15}.  Parameters used in theory are similar to those in our previous work modeling this data set\cite{Gullans15}, but also accounting for photoemission during lead tunneling: $g_c/2\pi = 80$~MHz, $\omega_c/2\pi =7.88$~GHz, $\kappa/2\pi=2.6$~MHz, $t_c=40$~$\mu$eV, $\Gamma_{L,R}/2\pi = 15$~GHz, $\gamma_d(0)/2\pi=3~$GHz, $c_n=4\, 000$~m/s,~$c_b=5\, 000$~m/s, $r=0.1$, $\omega_0/2\pi=30~$GHz, $T=2~$K and $\alpha_{L,R}={2.5\times 10^{-5}}$.  We accounted for charge noise by convolving these results with a Gaussian as function of $\epsilon$ with an RMS value of $40~\mu$eV.  (b) Comparison of predicted photoemission rate $\kappa_{\rm out} n$ to experimental data from Ref.~\onlinecite{Stockklauser15} including background photoemission.  As opposed to (a), these data were taken in the regime $2 t_c \approx \hbar \omega_c$: $g_c/2\pi =11$~MHz, $\omega_c/2\pi = 6.85$~GHz, $\kappa/2\pi=3.3$~MHz, $\kappa_{\rm out}/\kappa=0.5$, $t_c=13$~$\mu$eV, $\Gamma_{R,L}/2\pi=0.5$~GHz, $\gamma_d(0)/2\pi=400$~MHz, $c_b=4\, 000$~m/s,  $\omega_0/2\pi=25~$GHz $T=100$~mK and $\alpha_{L,R}={2.5\times 10^{-5}}$.  }
\label{fig:exp}
\end{center}
\end{figure}

To more quantitatively test our model for gain and photoemission, we directly compare our predictions to available experimental data from cavity coupled DQDs in InAs nanowires\cite{Liu15} and GaAs 2DEGs.\cite{Stockklauser15}
In the case of the comparison to measurements on an InAs nanowire DQD in Fig.~\ref{fig:exp}(a), we find that the TLS approximation is unable to account for the observed gain profile.  Note that by TLS approximation we mean that we only include the terms proportional to $r_c$ in Eq.~(\ref{eqn:chi}) and Eq.~(\ref{eqn:adaga2}) when computing $G$ and $n$.  The two possible extensions to the model we consider are (i) including the phonon-assisted emission and absorption processes and (ii) the photoemssion during lead tunneling.
  The phonon-assisted processes are accounted for by including the remaining terms in Eq.~(\ref{eqn:chi}) and Eq.~(\ref{eqn:adaga2}), with lead emission included via Eqs.~(\ref{eqn:lead1})--(\ref{eqn:lead2}).  
  In our previous work we only accounted for the addition of the phonon sideband and were able to obtain quantitative agreement with the data by fitting the electron-phonon coupling strength and the temperature.\cite{Gullans15,Muller17}   Here we also take into account the lead emission process and find that it contributes to the gain at positive detuning by an amount that is  comparable to the contribution from the phonon sideband.  However, the strong loss feature at negative detunings can not be explained by either lead emission (which always leads to gain) or the TLS approximation, therefore, we conclude that the phonon sideband and lead emission process contribute comparable amounts to the gain in these devices.  The large contribution from lead emission is consistent with the fact that these DQDs were operated in a regime of large lead tunnel rates ($\Gamma_{L,R}/2\pi \approx 15~$GHz).  As we discuss in Sec.~\ref{sec:phonon}, more direct signatures of the phonon sideband can be obtained by reducing the tunnel coupling to the leads.  

In Fig.~\ref{fig:exp}(b) we compare our theory to the measured photon emission rate $\kappa_{\rm out} n$ in a GaAs 2DEG DQD.  Here $\kappa_{\rm out}$ is the contribution to the total cavity decay rate $\kappa \approx \kappa_{\rm out}+\kappa_{\rm int}$ due to external damping out of the cavity and $\kappa_{\rm int}$ is the contribution from internal losses.     
In contrast to the InAs nanowire DQD, we find that these measurements largely agree with the TLS approximation with small deviations due to the phonon sideband and lead tunneling.  The most dominant deviations from the TLS approximation are seen in the tails in the photon emission rate at large $\epsilon$.  However, even in this regime,  the dominant correction to the TLS approximation arises from photoemission during lead tunneling.  The phonon sideband provides a larger correction to the photon emission rate near zero detuning when the DQD is approximately on resonance with the cavity.  This can be easily understood because the contribution to the photon emission from the phonon sideband becomes directly proportional to the temperature for GaAs 2DEG DQDs
\be \label{eqn:phoncont}
\lim_{\omega_d \to \omega_c} \mean{a^\dagger a}_{\rm ph} \propto  \frac{\Delta^{\alpha-1}}{\Delta^2+\eta^2} k_B T \approx k_B T.
\ee  
  This identity readily follows from Eq.~(\ref{eqn:game}) and Eq.~(\ref{eqn:adaga2}) with $\alpha=3$.  From this expression we see that measuring the photon emission as a function of temperature would better isolate the contribution from the phonon sideband.  Similarly the lead processes could be isolated and more carefully analyzed by measuring the photon emission rate for transport through a single quantum dot.  

In the case of Si 2DEG DQDs, we predict that the phonon sideband will have a weaker contribution to the photon emission rate on resonance 
\be 
\lim_{\eta \to0,\omega_d \to \omega_c} \mean{a^\dagger a}_{\rm ph}^{\rm Si} \propto \Delta^2 k_B T\to 0,
\ee
independent of temperature.  In this case, the effects of the phonon-assisted emission would primarily show up in tail  of the gain or photoemission rate at large $\epsilon$, as seen in Fig.~\ref{fig:gain}(f) and Fig.~\ref{fig:aa}(c).  These predictions for  2DEGs should be contrasted with the InAs nanowire, in which case, Eq.~(\ref{eqn:phoncont}) diverges as the renormalization parameter $ \eta$ (arising from the finite lifetime of the phonons and DQD excited state) is taken to zero.

\section{Phonon Spectroscopy}
\label{sec:phonon}

In the previous section we showed that signatures of the electron-phonon coupling appear in the cavity gain and photoemission rate.   We now show that the coupling to phonons has the additional effect of renormalizing the cavity frequency, which can be directly probed in the phase response to an input driving field [see Eq.~(\ref{eqn:phi})].  This effect has a similar origin as the dispersive shift of the cavity induced by the bare coupling to the DQD, but arises at 4th order in perturbation theory (2nd order in $g_c$ and 2nd order in the electron-phonon coupling) instead of 2nd order in $g_c$.

For the effective Hamiltonian in Eq.~(\ref{eqn:H}) the average shift in the cavity frequency $\delta_R$ in the NESS at zero temperature can be decomposed as
\begin{align}
\delta_R &= p_e (\delta_e + \delta_a) +  (p_e+p_g) \delta_\phi, \\
\delta_e &= \frac{16 g_c^2 t_c^2 \epsilon^2}{\omega_d^4 \omega_c^2}  \mathcal{P} \int  \frac{d \omega}{\omega^2+\eta^2} \frac{ (\omega+\omega_c)^2 J(\omega)}{\omega -\Delta+i 0^+}, \\
\delta_a &= \frac{16 g_c^2 t_c^2 \epsilon^2}{\omega_d^4 \omega_c^2}  \mathcal{P} \int  \frac{d \omega}{\omega^2+\eta^2} \frac{ (\omega-\omega_c)^2 J(\omega)}{\omega - (\omega_d +\omega_c)+i 0^+}, \\
\delta_\phi & = \frac{64 g_c^2 t_c^2 \sin^2 \theta}{(\Delta^2+\eta^2)(\omega_d+\omega_c)^2}  \mathcal{P} \int d\omega  \frac{J(\omega)}{\omega - \omega_c + i 0^+},
\end{align}
where $\mathcal{P}$ denotes the principal value.  These integrals are all UV convergent because of the exponential cutoff in $J(\omega)$ at large frequencies in our model, while the infrared divergences are regularized similar to Eqs.~(\ref{eqn:game})--(\ref{eqn:deph}).

\begin{figure}[t]
\begin{center}
\includegraphics[width = .49 \textwidth]{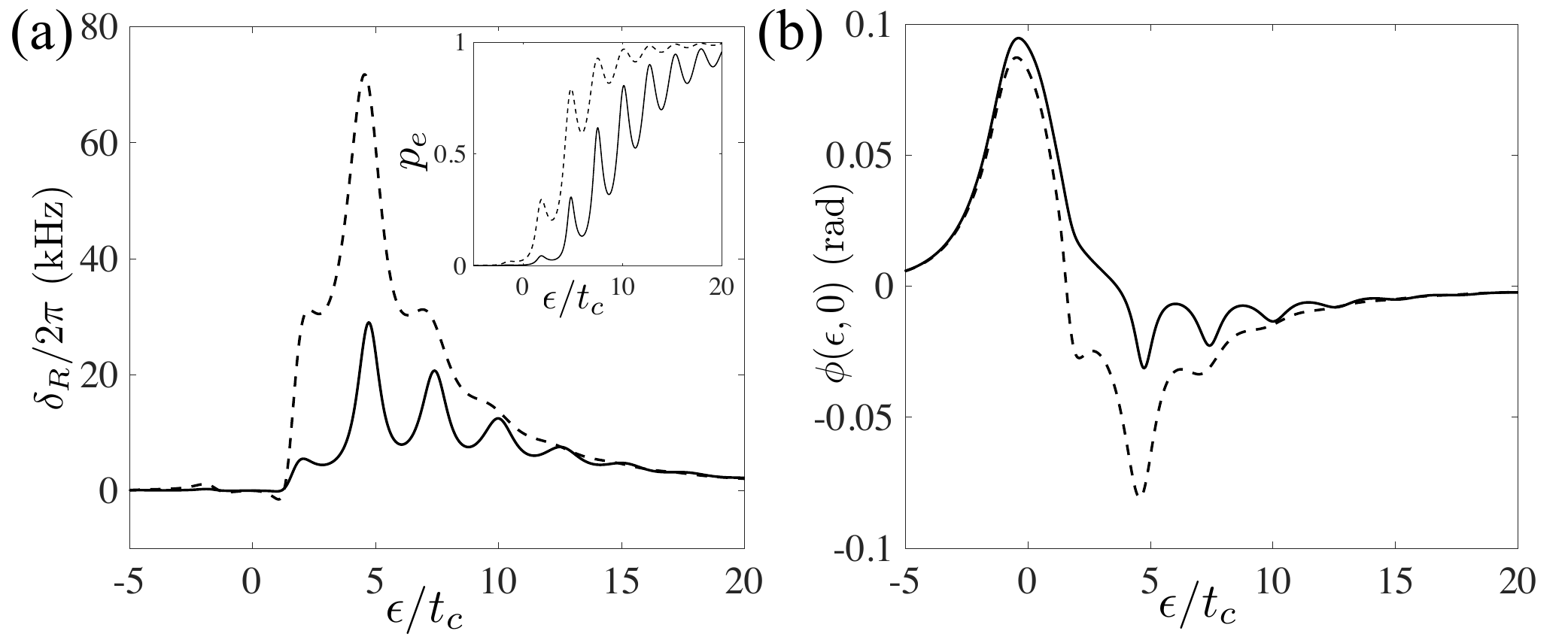}
\caption{(a) Renormalization of the cavity frequency due to electron-phonon coupling.  We took the spectral density for InAs nanowire DQD $J_{\rm nw}(\omega)$ defined in Eq.~(\ref{eqn:Jnw}).  Dashed and solid correspond to $\Gamma_R/2\pi=\Gamma_L/2\pi=1$~GHz and $\Gamma_R/2\pi=\Gamma_L/2\pi=100$~MHz, respectively. Other parameters were taken as $g_c/2\pi=40$~MHz, $\kappa/2\pi=2~$MHz, $\omega_c/2\pi=8~$GHz, $2t_c/h=16~$GHz, $\gamma_d(0)/2\pi$=2~GHz $c_n=2\, 100$~m/s, $d=124$ nm, $\omega_0/2\pi=130~$GHz, and $r=4\times 10^{-3}$. (inset) Population in excited state in the NESS. The visibility of the oscillations in $p_e$ decrease as the tunnel coupling to the leads is increased. (b) Phase response to an external drive field on resonance with the bare cavity frequency. }
\label{fig:phon}
\end{center}
\end{figure}

Similar to the gain enhancement we found in Sec.~\ref{sec:photon}, the term proportional to $\delta_e$ leads to a large shift in the cavity frequency for positive detunings due to the large population in the excited state.  These shifts can be used as a sensitive probe of the phonon spectral density.  In particular, as seen in Eqs.~(\ref{eqn:Jnw})--(\ref{eqn:J2d}), $J(\omega)$ generally exhibits oscillations in frequency with a period set by $c_n/d$.  These oscillations in $J(\omega)$ lead directly to oscillations in the decay rate $\gamma_d$.  The physical origin of the oscillations in $\gamma_d$ is the interference between phonon emission events on the two dots, which leads to either superradiance (anti-node on each dot) or subradiance (node on each dot).\cite{Brandes05}  These effects show up directly in the current through the dot and have been observed in several experiments and DQD material systems.\cite{Weber10,Roulleau11}  In the case of nanowire DQDs, there is still some remaining controversy over whether the oscillations in the current arise from the superradiant effect discussed here or from higher-order transverse phonon modes in the nanowire.

The periodic modulation in $\gamma_d$ has a strong effect on the population in $\ket{e}$ in the NESS [see inset to Fig.~\ref{fig:phon}(a)].  Most notably, when the tunneling rate off of the DQD $\Gamma_R$ is less than or comparable to the average value of $\gamma_d$, the excited state population will be highly sensitive to small changes in $\gamma_d$ such that the oscillations in $J(\omega)$ are directly mapped onto the excited state population.  The effect of this periodic modulation on the cavity frequency and phase response is shown in Fig.~\ref{fig:phon}(a) and Fig.~\ref{fig:phon}(b), respectively, for two different values of $\Gamma_R$.  We can clearly see that for small values of $\Gamma_R$ the cavity frequency and phase response show clear oscillations, which are directly correlated with the oscillations in $J(\omega)$.  As a result, we see that the cavity frequency serves as a highly sensitive probe of the electron-phonon coupling in this system.  This effect was recently directly observed in a suspended InAs nanowire DQD.\cite{Hartke17}

\section{Conclusions}

We investigated the role of electron-phonon coupling in the charge-photon dynamics of cavity-coupled DQDs.  By making direct comparisons between three different quantum dot material systems we were able to  better understand similarities and differences in their charge-photon dynamics.  Most notably we find that electron-phonon coupling has the strongest signatures in InAs nanowire DQDs due to the enhanced 1D phonon density of states at low-energies.  In GaAs 2DEG DQDs the phonon sideband should be directly observable by investigating the temperature dependence of the photoemission rate when the DQD is directly on resonance with the cavity.  In contrast, in Si 2DEG DQDs the phonon sideband plays a much weaker role due to the absence of piezoelectric coupling.  Nevertheless, its contributions should be observable in photoemission and gain profile when the DQD TLS has a much higher energy than the cavity.  We expect that this work will help guide future efforts in finding optimal operating points for quantum dot masers and performing detailed microwave spectroscopy of electron-phonon interactions.  Such efforts will aid in the realization of large scale quantum networks with semiconductor spin qubits and superconducting cavities.

\begin{acknowledgements}
We thank T. R. Hartke, D. A. Huse, and Y.-Y. Liu for helpful discussions. We also thank A. Stockklauser and A. Wallraff for helpful discussions and for sharing their experimental data. Research was supported by the Gordon and Betty Moore Foundation’s EPiQS Initiative through Grant GBMF4535, with partial support from the National Science Foundation through DMR-1409556 and DMR-1420541. Devices were fabricated in the Princeton University Quantum Device Nanofabrication Laboratory.
\end{acknowledgements}

\appendix

\section{Phonon Spectral Density} 

In this Appendix we use a toy model to derive  the low-energy DQD phonon spectral densities used in this work. 
Neglecting electron interaction effects, we can express the matrix elements for the electron-phonon interactions in the DQD using single-particle wavefunctions \cite{Brandes05}
\begin{align}
\lambda_{\bq \nu} &= \bra{L} V_{\bq\nu} \ket{L} - \bra{R} V_{\bq\nu} \ket{R} \\ \nonumber
&=\int d \bm{x}  (\abs{\phi_\ell(\bm{x})}^2-\abs{\phi_r(\bm{x})}^2) V_{q\nu}(\bm{x}) e^{i \bq \cdot \bm{x}} \\ \nonumber
&=2i\sin(q_z d/2) e^{-i q_z d/2} M_{\bq\nu},\\ 
M_{\bq\nu}&=\int d\bm{x} \abs{\phi(\bm{x})}^2  V_{\bq\nu}(\bm{x}) e^{i \bq\cdot \bm{x}},
\end{align}
where $\ket{L/R}$ are the single-particle states of the left/right dot, $V_{\bq\nu}$ is the interaction potential for the phonon mode with wavevector $\bq$  and branch $\nu$,  $d$ is the spacing between the dots, the dots are taken to lie along the $z$-axis, and $\phi_{\ell(r)}=\phi$ are the envelope wavefunctions for the electrons in the dots.  We  approximate the envelope by  a spherical  Gaussians $\phi(\bm{x}) \propto  e^{-|\bm{x}|^2/2 a^2}$.   The electron-phonon interaction potential can be broken up into a contribution from the deformation potential  and the piezoelectric potential.\cite{Mahan}   Expanding $V_{q\nu}$ and performing the integral over space gives the form
\be
M_{\bq\nu}= \sqrt{\frac{1}{2 M\, \omega_\nu(\bq) }}\,\, (i |\bq| \beta_\nu   +  \Xi_\nu) \, e^{-a^2 |\bq|^2/4},
\ee
where $M$ is the average mass of the unit cell and $\beta_\nu$ and $\Xi_\nu$ are deformation and piezoelectic constants.  The expression for the phonon spectral density in 1D $J_{\rm nw}(\nu)$ given in Eq.~(\ref{eqn:Jnw}) follows directly from these results.  In 3D one has to perform an additional integral over the angular variables to arrive at $J_\alpha(\nu)$ given in Eq.~(\ref{eqn:J2d}). 

\bibliographystyle{apsrev-nourl-title-PRX}
\bibliography{DDMaser}

\end{document}